\documentclass[eqsecnum,amsmath,preprintnumbers,nofootinbib,aps]{revtex4} 
\usepackage{epsfig}
\usepackage{amssymb}
\usepackage{subfigure}
\usepackage{color}
\usepackage{hyperref}
\usepackage{graphicx}
\usepackage{enumerate}
\usepackage[countmax]{subfloat}
\usepackage{MnSymbol}

\DeclareMathAlphabet{\mathpzc}{OT1}{pzc}{m}{it}

\def\beq{\begin{equation}}
\def\eeq{\end{equation}}

\def\bea{\arraycolsep .1em \begin{eqnarray}}
\def\eea{\end{eqnarray}}
\def\Tr{{\rm Tr}}
\def\tr{{\rm tr}}

\def\g{\gamma}

\def\eq#1{(\ref{#1})}

\def\s0#1#2{\mbox{\small{$ \frac{#1}{#2} $}}}
\def\0#1#2{\frac{#1}{#2}}

\def\grgl{\:\hbox to -0.2pt{\lower2.5pt\hbox{$\sim$}\hss}{\raise3pt\hbox{$>$}}\:}
\def\klgl{\:\hbox to -0.2pt{\lower2.5pt\hbox{$\sim$}\hss}{\raise3pt\hbox{$<$}}\:}

\begin{document}

 \title{Renormalisation of Newton's constant}
\author{Kevin Falls}

\affiliation{Instit\"{u}t f\"{u}r Theoretische Physik, Universit\"{a}t Heidelberg, Philosophenweg 12, 69120 Heidelberg, Germany}

\begin{abstract}
The problem of obtaining a gauge independent beta function for Newton's constant is addressed. 
By a specific parameterisation of metric fluctuations a gauge independent functional integral is constructed for the semiclassical theory around an arbitrary Einstein space. The effective action then has the property that only physical polarisations of the graviton contribute, while all other modes cancel with the functional measure. We are then able to compute a gauge independent beta function for Newton's constant in $d$-dimensions to one-loop order.
No Landau pole is present provided $N_g < 18$, where $N_g = d(d-3)/2$ is the number of polarisations of the graviton. While adding a large number of matter fields can change this picture, the absence of a pole persists for the particle content of the standard model in four spacetime dimensions.

\end{abstract}

\maketitle
\date{\today}
\maketitle
\newpage
\tableofcontents

\newpage

\section{Introduction}

In quantum field theory it is well known that coupling constants become functions of the energy scales entering the renormalisation process. In turn this implies a modification of the classical scaling properties of a theory \cite{Wilson:1973jj}. Such energy dependence of a coupling $a$ is encoded in its beta functions $\beta_{a} = \mu \frac{\partial}{\partial \mu} a$, where $\mu$ is the renormalisation scale. In  quantum chromodynamics (QCD) this non-trivial scaling can be observed in the one-loop beta function \cite{Gross:1973id,Politzer:1973fx},
\beq \label{betaQCD}
\beta_{\alpha_s} = - \left(11 - \frac{2}{3} N_f \right) \frac{\alpha_s^2}{2 \pi}\,,
\eeq
where $\alpha_s$ is the strong coupling and $N_f$ denotes the number of flavours. For $N_f \leq 16$ this equation describes the weakening of the strong force as the energy scale $\mu$ is increased. On the other hand if we have $N_f > 16$ the coupling constant will diverge at a finite energy. In the former case the theory is said to be asymptotically free and is well defined for all scales. In the later case the coupling has a Landau pole and the theory can only be considered as an effective one, at energies below the pole.

When considering gravity as a quantum field theory it is natural to ask whether there exists an analog to \eq{betaQCD} for Newton's constant $G_N$. 
 In four spacetime dimensions 
 Newton's constant is dimensionful which implies that the structure of the beta function differs from that of QCD, where $\alpha_s$ is dimensionless. In particular, for $d$ spacetime dimensions one should work with the dimensionless coupling $G= \mu^{d-2} G_N$. The analog of \eq{betaQCD} will then take the form
\beq \label{betab}
\beta_G = (d-2) G - b \, G^2 \,,
\eeq
for some constant $b$. Whereas the coefficient of $G$ just reflects the dimensionality of $G_N$ the coefficient $b$ is a quantum correction \footnote{b is proportional to $\hbar$, although here we work in natural units $\hbar = 1 = c$.}. Clearly $b$ plays the same role as $11 - \frac{2}{3} N_f $ in the QCD; if $b$ is positive then $G$ will remain finite for all scales, but if $b$ is negative $G$ will blow up at a finite energy. Thus, although non-perturbative effects may change the naive conclusion, the sign of $b$ is a strong indiction of whether quantum gravity should be replaced by some new theory at a finite energy scale, or whether gravity may itself constitute a perfectly valid quantum field theory. The latter scenario, dubbed asymptotic safety by Weinberg \cite{Weinberg:1980gg}, would involve a fixed point for $G$ where $\beta_G =0$ for some positive value $G=G_*$. At the level of the one-loop beta function \eq{betab} this is found at $G_* = \frac{1}{b}(d-2)$ and its existence only depends on the sign of $b$. However, one must go beyond one-loop in order to be sure that such a fixed point exists non-perturbatively. 

Although $\beta_G$ has been calculated previously, in these calculations the value of $b$ has been found to depend on the gauge \cite{Falkenberg:1996bq, Lauscher:2001ya, Litim:2003vp,Percacci:2015wwa} and parameterisation \cite{Nink:2014yya,Percacci:2015wwa} (an exception are calculations with the geometrical effective action \cite{Donkin:2012ud}, which are gauge independent although a particular parameterisation of the fields is used). 
Hence these calculations will generically include unphysical degrees of freedom coming from the gauge fixing action and may not be trust worthy. This issue stems from the fact that calculations have to be performed off-shell and therefore generally depend on both the gauge and the parameterisation. When working on-shell such dependencies vanish. On the other hand the Einstein equations dictate that the scalar curvature is given by
\beq \label{Ronshell}
R =  \frac{2d}{d-2} \bar{\lambda}\,,
\eeq
 where $\bar{\lambda}$ is the cosmological constant. As a consequence one cannot disentangle the renormalisation of Newton's constant from the vacuum energy, forcing us to work off-shell to determine $b$. Thus, it is clear that by working off-shell, in a gauge dependent setup, no definitive conclusion can be reached as to the sign of $b$. On the other hand explicit non-perturbative calculations both in four and higher dimensions indicate that a fixed point for $G$ exists \cite{Souma:1999at,  Reuter:2001ag, Lauscher:2001ya, Percacci:2002ie, Litim:2003vp,Fischer:2006fz,Donkin:2012ud,Christiansen:2014raa, Becker:2014qya, Falls:2014zba}, despite calculations being generally gauge dependent. Hence, although evidence is strong that the coupling $G$ does not blow up for some finite energy scale, it is also questionable since the dependence on the gauge parameters implies that unphysical contributions are still present.

In this paper we shall obtain gauge independent results by disentangling physical degrees of freedom at the level of the functional integral. In particular we shall obtain a semiclassical approximation to the 
$d$-dimensional functional integral in quantum gravity on an arbitrary Einstein space independent of the gauge {\it without} fixing $R$ by the on-shell condition \eq{Ronshell}. This functional integral has the property that it only receives 
contributions from physical fluctuations of the metric that survive on-shell. This leads us to gauge independent beta function for the gravitational constant.  
We will then see that the {\it sign} of $b$ can be universally determined by a unique factor
\beq \label{betaintro}
b \propto  \frac{2}{3} (18-N_g)\,,\,\,\,\,\,\,\,\,\, {\rm with }  \,\,\,\,\,\,\,\,\, N_g = \frac{d(d-3)}{2}\, ,
\eeq 
where $N_g$ is the number of dynamical degrees of the metric in general relativity. Thus one finds that $b$ is positive in $d=4$ dimensions but becomes negative for integer dimension $d\geq 8$. This result is for pure gravity but is also easily generalised when matter is included.

At this point we should take a moment to clarify the meaning of \eq{betaintro}. Although one may like to think of $G$ being related to an observable, this may not be the case. In particular, at least perturbatively, there is no universal definition of a running Newton's constant for e.g. scattering processes  \cite{Anber:2011ut}.
What \eq{betaintro} does tell us is whether or not the ultra-violet (UV) cutoff $\Lambda$ can be removed and hence whether a continuum limit exists. For $N_g <18$ one can remove $\Lambda \to \infty$ without any problems. However, for $N_g >18$ one may show that for the bare Newton's constant to be positive, and hence for the functional integral to exist, $\Lambda$ must remain finite.
Thus the beta function implies the existence  (non-existence) of a continuum limit for $N_g < 18$ ($N_g >18$). The universal and gauge independent result \eq{betaintro} therefore provides evidence for the continuum limit of quantum gravity in four spacetime dimensions. 

The rest of this paper is as follows. In section~\ref{Measure} we find the form of the functional measure for the gauge fixed functional integral over geometries and use it to determine the on-shell functional integral for the semi-classical theory.
This generalises the result of \cite{Mazur:1989by} to all dimensions $d >2$. Turning to section~\ref{SemiClass}, we then confront the problem of obtaining gauge independent results off-shell finding a parameterisation of metric fluctuations which achieve this end given in section~\ref{Gindep}. We are then able to write down a gauge independent one-loop effective action. In section~\ref{Ginvar} the gauge independence is linked to the gauge invariance of the quadratic action achieved while using our specific parameterisation. Since our gauge independent result is achieved using a transverse-traceless decomposition in section~\ref{deDonder} we also calculate the effective action in de Donder gauge as a cross-check and find a form of the effective action in terms of unconstrained fields. Section~\ref{Low_energy} is devoted to a calculation of quantum corrections to the Newtonian potential showing that the result is independent of the parameterisation.
 We then use our parameterisation to derive a semiclassical renormalisation group equation for a scale dependent action $\Gamma_k$ in section~\ref{flow}. In section~\ref{HeatKernels} we evaluate traces appearing in the flow equation using the early time heat kernel expansion and access their universal content. We may then observe the universal factor  \eq{betaintro} while also reproducing the curvature squared counter term found in \cite{Christensen:1979iy}. Then in section~\ref{betaGsec} we give the explicit form of  the beta function $\beta_G$. We first discuss the case near two dimensions and the limit $d \to2$. Then we consider the general case in $d$ dimensions and discuss in more detail the physical origin of the universal factor \eq{betaintro}. In section~\ref{vacuum} we find the form of the divergent counter term for the vacuum energy and note that it vanishes in $d=3$ dimensions. In section~\ref{continuum} we derive a bound on the UV scale $\Lambda$ of the theory in higher dimensions and show no bound exists for $N_g < 18$. In Section~\ref{matter} we present the beta function with the inclusion of matter.  We end with our conclusions in section~\ref{conclude}.

\section{The on-shell functional integral}  \label{Measure}

General relativity in $d$ dimensions describes $(d-3) d/2$ dynamical degrees of freedom coming from the $d(d+1)/2$ components of the metric, which are determined by Einstein equations, minus $d$ constraints and $d$ diffeomorphisms. In the quantum theory these degrees of freedom correspond to the $N_g$ polarisations of the graviton, schematically one has
\beq \label{Ng}
N_g\,\,\, = \,\,\, \underbrace{d(d+1)/2}_{\rm metric}  \underbrace{-\,\,\,\, \,\,\,\,d}_{\rm diffeomorphisms}  \,\,\,\,   \underbrace{-\,\,\,\, \,\,\,\, d}_{\rm  constraints} =\,\,\, d(d-3)/2 \,.
\eeq
Thus one should expect that the additional $2d$ unphysical degrees of freedom are removed when physical quantities are computed. At the level of the functional integral such cancelations occur between metric fluctuations and the functional measure. One way to ensure this cancelation occurs is to turn off any external source terms which violate parameterisation invariance, equivalently, when working with the effective action, this can be achieved by working on-shell. In this section we will determine the form of the semi-classical functional integral by expanding around a solution to the equations of motion to ensure gauge and parameterisation independence. This we do as an intermediate step to gain physical intuition and see mathematically how the unphysical states are removed when computing on-shell quantities.  

The first step in our calculation is to determine the form of the functional measure for quantum gravity. We assume a bare action for Eunclidean quantum gravity of the Einstein-Hilbert form
\beq \label{action}
S_{\rm grav}[\g_{\mu\nu}] =  \frac{1}{16 \pi G_b} \int d^d x \sqrt{\g}  \left[\, 2 \bar{\lambda}_b -  R(\g_{\mu\nu})\, \right]  + S_{\rm gf}[\g_{\mu\nu}]\,,
\eeq
where $S_{\rm gf}$ is the gauge fixing action. Here $\g_{\mu\nu}$ is the metric tensor with determinant $\g = \det \g_{\mu\nu}$  and $R(\g_{\mu\nu})$ denotes the Ricci scalar. The action depends on two parameters $G_b$ and $\bar{\lambda}_b$ which denote the bare Newton's constant and bare cosmological constant respectively. The functional integral is then given by
\beq \label{Z}
\mathcal{Z} = \int \mathcal{D} \g_{\mu\nu} (\det{\mathcal{Q}})\, e^{-S_{\rm grav}[\g_{\mu\nu}]}  
\eeq
where $\det{\mathcal{Q}}$ is the determinant of the Faddeev-Popov operator. 

Here we shall evaluate the functional integral \eq{Z} in the semiclassical approximation. To this end we first consider parameterisations of metric fluctuations $h_{\mu\nu}$ in terms of a background field $g_{\mu\nu}$ which satisfies the classical equations of motion \footnote{ Here we adopt the notation that all curvatures and covariant derivatives are with respect to $g_{\mu\nu}$ unless indicated otherwise, such that $R \equiv R(g_{\mu\nu})$ etc. and $\nabla_\mu = \nabla_\mu(g_{\mu\nu}) $ is the covariant derivative with respect to $g_{\mu\nu}$, and indices are lowered and raised with $g_{\mu\nu}$.}
\beq \label{onshell}
R_{\mu\nu}  = \frac{2}{d-2} \bar{\lambda}_b \, g_{\mu\nu} \,.
\eeq
Two such parameterisations are the linear parameterisation
\beq \label{linear}
\g_{\mu\nu} = g_{\mu\nu} + h_{\mu\nu}   \,,
\eeq
and the exponential parameterisation
\beq \label{expo}
\g_{\mu\nu} = g_{\mu\lambda} [e^h]^{\lambda}_\nu = g_{\mu\nu} + h_{\mu\nu} + \frac{1}{2} h_{\mu}^\lambda h_{\lambda \nu} + ... \,,
\eeq
where in the later case $h$ is a symmetric matrix with components $[h]^\mu_\nu \equiv  h^\mu_\nu$. 
An advantage of the exponential parameterisation \eq{expo} is that one may single out the conformal factor of the metric by  
\beq \label{conffactor}
\g_{\mu\nu} = e^{\frac{\bar{\phi}}{d}} g_{\mu\lambda} [e^{\hat{h}}]^{\lambda}_\nu \equiv e^{\frac{\bar{\phi}}{d}} \hat{\g}_{\mu\nu}
\eeq
where $\bar{\phi} = h^\mu_\mu$ is the trace of $h$ and $\hat{h}$ is the trace free part. From here it follows that the determinant of $\hat{\g}$ is fixed to the background one
\beq \label{detg}
\hat{\g} = g
\eeq
and thus $e^{\frac{\bar{\phi}}{d}}$ is identified as the conformal factor of $\g_{\mu\nu}$ with respect to the background metric $g_{\mu\nu}$.

We shall need to expand the action to second order in the fluctuation around the on-shell metric $g_{\mu\nu}$.
 To ensure the resulting hessians are invertible we must also fix a gauge, the choice of which should fall out of physical quantities. Here we choose a class of background field gauges, linear in the fluctuation $h^\mu_\nu$ for either parameterisation, 
 \beq \label{gf}
S_{\rm gf} = \frac{1}{ 32 \pi G_b \alpha}  \int d^d x \sqrt{g}  g^{\mu\nu} F_{\mu} F_{\nu} \,,\,\,\,\,\,\,\, \,\,\,\,\,\,\,\,\,\,\,\,\,\,      F_\mu =\nabla_\lambda h^\lambda_\mu - \frac{1 + \rho}{d} \nabla_\mu h^\lambda_\lambda \,.
 \eeq
 We now follow the steps out lined in \cite{Benedetti:2011ct} to decompose the fluctuation $h_{\mu\nu}$ such that it becomes manifest that the gauge fixing action only depends on $d$ independent fields matching the number of diffeomorphisms. The first step is to adopt the transverse-traceless decomposition of the fluctuations in terms of differentially constrained fields \cite{York:1973ia},
\bea \label{TT}
h_{\mu\nu} &= &h^\upvdash_{\mu\nu} + \bar{\phi}  \frac{1}{d} g_{\mu\nu}+ \nabla_{\nu} \xi_{\mu} +\nabla_{\mu} \xi_{\nu} + \nabla_{\mu}\nabla_{\nu}  \bar{\psi} - \frac{1}{d} g_{\mu\nu} \nabla^2   \bar{\psi}\,,\\[2ex]
&& \,\,\, h^{\upvdash \mu}_{\mu}= 0\,, \,\,\,\,\,\,\,\,\,\,\,\,\nabla_\mu h^{\upvdash \mu}_\nu = 0 \,,\,\,\,\,\,\,\,\,\,\,\,\nabla_{\mu}\xi^{\mu} = 0 \,. \nonumber
\eea
One observes that $\xi_\mu$ takes the form of a transverse diffeomorphism of the metric to linear order.
Thus we can identify $\xi_\mu$ as $d-1$ of the unphysical fields corresponding to such diffeomorphisms. 
We then further redefine the scalar fields $\{\bar{\phi},\bar{\psi} \} \to \{ \phi, \psi \}$ as
\beq \label{DBredef}
\bar{\phi}= \phi + \nabla^2 \bar{\psi}\,,  \,\,\,\,\,\,\,  \bar{\psi} = \psi + \frac{\rho}{(d-1-\rho) \nabla^2 + R} \phi \,,
\eeq
for which it becomes manifest that the gauge fixing action \eq{gf} only depends on the transverse vector $\xi_\mu$ and the scalar $\psi$.
Thus we see that $\psi$ represents the additional longitudinal diffeomorphism. One should then expect that the integral of these fields in \eq{Z} should be cancelled by the functional measure.
The field redefinition \eq{TT} leads to the following Jacobians in the functional measure 
\beq \label{Jh}
J_h = \sqrt{{\det}_0[\Delta]}  \sqrt{{\det}_0\left[\Delta_0\right]}  \sqrt{{\det}_{1T}\left[\Delta_1\right]} \,,
\eeq
whereas the redefinition \eq{DBredef} has a trivial Jacobian.
The non-trivial Jacobian \eq{Jh} is given by determinants of  the differential operators,
\bea \label{Deltas}
\Delta \varphi &=&  -\nabla^2\varphi \nonumber \\
\Delta_0 \varphi &=&    \left(-\nabla^2- \frac{R}{d-1}\right) \varphi \\
\Delta_1 \varphi_\mu&=&    \left(-\nabla^2 \delta_\mu^\nu- R_{\mu}\,^\nu \right) \varphi_\nu \nonumber \,,
\eea
acting on scalars and vectors as indicated. The subscripts $1T$ and $0$ specify that the determinants are evaluated from transverse vectors and scalars.
 It is important to note that the constant mode of the scalar $\psi$ should be left out of the functional measure since it cannot contribute to the fluctuation $h_{\mu\nu}$. Accordingly the the constant modes in the scalar Jacobians should also be removed. Similarly one should leave out the Killing vectors $\nabla_{\nu} \xi_{\mu} + \nabla_{\mu} \xi_{\nu} = 0$ from the vectors $\xi^\mu$ and the transverse vector Jacobian $\sqrt{{\det}\left[\Delta_1\right]}$, this corresponds to the zero mode of $\Delta_1$. 

Next we turn to the Faddeev-Popov determinant. Here we will exploit the liberty to write the determinant as $\det \mathcal{Q} =  \sqrt{\det \mathcal{Q}^2}$ such that the ghosts become second order in derivatives at the price of including a set of commutative ghosts fields $B^\mu$ in addition to the anticommuting ghosts $\bar{C}_\mu$ and $C_\mu$. We then decompose the ghosts into transverse and longitudinal fields
\beq \label{ghdecom}
B_\mu = B_\mu^T +\nabla_\mu B^L\,,\,\,\, C_\mu = C_\mu^T +\nabla_\mu C^L\,,   \,\,\,\, \bar{C}_\mu = \bar{C}_\mu^T +\nabla_\mu \bar{C}^L \,,
\eeq
where the superscript $T$ indicates that the field is transverse i.e. $\nabla^\mu B^T_\mu = 0$. 
This transformation then leads to a Jacobian $J_{\rm gh} = \frac{1}{\sqrt{{\det}_0[\Delta]}}$ which cancels with the corresponding factor in the metric fluctuation Jacobian $J_h$ given by \eq{Jh} leaving just two Jacobians,
\beq \label{Js}
J_0 = \sqrt{{\det}_0 \left[\Delta_0\right]}\,, \,\,\,\,\,\,\,  J_1= \sqrt{{\det}_{1T} \left[ \Delta_1 \right] }\,.
\eeq
It is these Jacobians that remove the $d$ degrees of freedom corresponding to the constraints whereas the ghosts should cancel the integral over diffeomorphisms (i.e. $\xi_\mu$ and $\psi$). 
Due to the decomposition \eq{ghdecom} the corresponding Faddeev-Popov determinants splits into two factors for the longitudinal and transverse parts,
\beq \label{Qoperators}
\det \mathcal{Q}_L =   \sqrt{{\det}_0[ \Delta_L^2 \Delta]} \,, \,\,\,\,\, \det \mathcal{Q}_T =   \sqrt{{\det}_{1T}\left[\Delta_1^2 \right]}
\eeq
where we define the gauge dependent differential operator $\Delta_L$ by
\beq \label{DeltaL}
\Delta_L \varphi =\left(-\nabla^2 - \frac{R}{d-1 - \rho } \right) \varphi \,,
\eeq
acting on scalar fields.
The path integral is then given by
\beq \label{Zdecom}
\mathcal{Z} = \int \mathcal{D} h^\upvdash_{\mu\nu}  \mathcal{D}\phi  \mathcal{D}\xi_\mu  \mathcal{D} \psi \, J_0\,  J_1 \det (\mathcal{Q}_L) \det(\mathcal{Q}_T) \, e^{-S_{\rm grav}[\phi ,h^\upvdash,\psi,\xi]}  \,.
\eeq

To obtain the semiclassical functional integral we must then expand the action to second order in the fluctuations of the fields $\{\phi ,h^\upvdash,\psi,\xi \}$ while taking into account the contributions from the measure. 
 The next stage is therefore to compute the second variation of the bare action and apply the on-shell condition \eq{onshell}. 
 
 First let's show that the choice of parameterisation, such as \eq{linear} or \eq{expo}, does not effect the on-shell hessians. Here we just assume that the metric is a function of the fluctuation $h_A(x)$ (with e.g. $A =\{\mu\nu\}$)  such that
 \beq \label{para}
 \g_{\mu\nu} = \g_{\mu\nu}(h_A(x)) \,\,\,\,  {\rm with }\,\,\, \g_{\mu\nu}(0) =  g_{\mu\nu} 
 \eeq 
 which agree only up to linear order in the fluctuation. The choices \eq{expo} and \eq{linear} are just two examples of such parameterisations.
 Introducing de Witt's condensed notation we write $i = \{ x, \mu \nu \}$ and $a = \{x ,A\}$ and take two functional derivatives with respect to the fluctuations $h_A(x) \equiv h^a$ expressed with derivatives with respect to the metric $ \g_{\mu\nu}(x) \equiv  \g^{i}$. Acting on e.g. $S[\gamma^i] = S_{\rm grav}$ we obtain,
 \beq \label{hderivatives}
 \frac{\delta^2 S[\g^k[h^a]]}{\delta h^a \delta h^b} = \frac{\delta}{\delta h^b}  \left( \frac{\delta \g^i}{\delta h^a}   \frac{\delta S[\g^i]}{\delta \g^i} \right)  = \frac{\delta^2 \g^i}{\delta h^a h^b}   \frac{\delta S[\g^i]}{\delta \g^i} + \frac{\delta \g^i}{\delta h^a} \frac{\delta \g^j}{\delta h^b}   \frac{\delta S[\g^i]}{\delta \g^i\delta \g^j} \,.
 \eeq
 on-shell the first term is zero since it is proportional to the equation of motion, while the second term only depends on $\g_{\mu\nu}$ to linear order of  in $h_{\mu\nu}$. 
This shows that by going on-shell the results cannot depend on which of parameterisations is used.

The on-shell hessians $S^{(2)}_{\rm grav}$  has the following components, for metric fluctuation fields $\xi_\mu$, $\psi$ and $h^{\upvdash}_{\mu\nu}$, 
  \beq \label{S2}
 16 \pi G_b \, S^{(2)}_{\xi\xi} =  \frac{1}{\alpha} \Delta_1^2\,,  \,\,\,\,\,\,\,\,\, 16 \pi G_b \, S^{(2)}_{\psi \psi} = \frac{(d-1 - \rho)^2}{\alpha d^2} \Delta_L^2 \Delta \,, \,\,\,\,\, 16 \pi G_b \, S^{(2)}_{h^\upvdash h^\upvdash}= \frac{1}{2} \Delta_2 \,,
 \eeq
 where 
 \beq
\frac{\delta^2 S_{\rm grav}}{ \delta \varphi^A(x) \varphi^B(y)} = \sqrt{g}\, S^{(2)}_{\varphi^A\varphi^B} \, \bold{1}_{AB} \, \delta(x-y)
\eeq
with $\varphi^A$ denoting the various fields and $\bold{1}_{AB}$ is the identity on the corresponding field space. 
In \eq{S2} the differential operator $\Delta_2$ is the Lichnerowicz Laplacian acting on two tensors as
\beq \label{Delta2}
\Delta_2 \varphi_{\mu\nu} = -\nabla^2 \varphi_{\mu\nu} - 2 R_{\mu}\,^\alpha\,_\nu\,^\beta \varphi_{\alpha \beta} \,.
\eeq
The field $\phi$, whose on-shell hessian is given by 
\beq \label{S2phih}
 16 \pi G_b \, S^{(2)}_{\phi\phi} = -\frac{(d-1)(d-2)}{2 d^2} \Delta_{0}\,,  
\eeq
needs extra attention since one observes that \eq{S2phih} 
is negative for positive eigenvalues of the operator $\Delta_0$. This
indicates that the naive Wick rotation of the functional integral is unbounded from below.
However, as pointed out by Mottola and Mazur \cite{Mazur:1989by} one should Wick rotate all modes with positive eigenvalue as $\phi \to i \phi$ whereas modes with negative eigenvalues should be Wick rotated trivially. This rule can be derived by considering the super-metric on the space of fluctuations $h_{\mu\nu}$.  Hence one can write the corresponding operator as
\beq \label{S2phi}
16 \pi G_b \, S^{(2)}_{\phi\phi} =  \frac{(d-1)(d-2)}{2 d^2} | \Delta_0 | \,,
\eeq 
to get a well defined Euclidean functional integral.

Expanding the action to second order in the fluctuation, and comparing $S^{(2)}_{\xi\xi}$ and $S^{(2)}_{\psi\psi}$ given in \eq{S2}  to  \eq{Qoperators}, one observers that the functional integrals over $\xi$ and $\psi$ cancel with the Faddeev-Popov determinants $\det \mathcal{Q}_L$ and $\det \mathcal{Q}_T$. All gauge dependence has therefore cancelled out. Furthermore comparing \eq{S2phi} with $J_0$ we see that  all modes of $\phi$ apart from the constant mode
\beq
\partial_\mu \phi_0 = 0\,,
\eeq 
are cancelled by the Jacobian $J_0$ given by \eq{Js} (here we assume that the constant mode is the only mode for which $\Delta_0$ has a negative eigenvalue, which is true at least for the $d$-sphere). This follows since, as we argued, the constant mode must be left out of $J_0$.
The only contributions that remain are  the  $ (d-2) (1+d)/2$ transverse-traceless fluctuations, the Jacobian $J_1$, which comprises $d-1$ negative degrees of freedom, and the constant mode $\phi_0$. Thus one is left with $(d-3)d/2$ local degrees of freedom corresponding to the graviton and one global degree of freedom corresponding to a constant rescaling of the metric. 
The functional integral then reduces to the form
\beq \label{Zfinal}
\mathcal{Z} = \int d\phi_0 \int \mathcal{D} h^\upvdash \sqrt{{\det}_{1T}\left[\Delta_1\right]} \exp \left[- S_{\rm grav}[g] -\frac{1}{32 \pi G_b} \int d^dx \sqrt{g} \left( h_{\mu\nu}^\upvdash \Delta_2 h^{\upvdash \mu\nu} - \frac{(d-1)(d-2)}{2 d^2}  \phi_0 \Delta_0 \phi_0 \right) \right] \,,
\eeq
this result is valid in all dimensions $d >2$ and for all parameterisations \eq{para} of the metric fluctuations. The remaining determinant can be expressed in terms of integral over auxiliary fields
\bea \label{J1}
J_1 = \sqrt{{\det}_{1T} \left[\Delta_1\right]} &=& \int \mathcal{D}\zeta_\mu \mathcal{D} c_\mu \mathcal{D} \bar{c}_\mu  \exp\left[  -\frac{1}{32 \pi G_b} \int d^dx \sqrt{g} \left( \bar{c}_\mu \Delta_1 c^\mu + \zeta_\mu \Delta_1 \zeta^\mu   \right)   \right] \nonumber \\ &\equiv&  \int \mathcal{D}\zeta_\mu \mathcal{D} c_\mu \mathcal{D} \bar{c}_\mu  e^{-S_{\rm aux}[ \bar{c}_\mu,  c_\mu , \zeta_\mu] }\,,
\eea
where $ \bar{c}_\mu,c_\mu$ are anticommuting and $\zeta_\mu$ is commuting and all are transverse. Here $S_{\rm aux}$ is the auxiliary action that combines with the gravity action in \eq{Zfinal} to form a modified bare action $S = S_{\rm grav} + S_{\rm aux}$.  The semiclassical functional integral \eq{Zfinal} generalises the $d=4$ result of \cite{Mazur:1989by}.  We note that, while the cancellation of the $d$ diffeomorphisms and one of the constrained degrees of freedom is explicit, the cancellation between the remaining $d-1$ unphysical transverse traceless fluctuations is implicit and implies that non-trivial cancelations between these modes and the Jacobian $J_1$ should occur. It is clear therefore that one should not leave out the Jacobians from the functional, as was done in \cite{Demmel:2014hla}, doing so implicitly includes $d$ unphysical degrees of freedom and fails to reproduce the one-loop functional integral \eq{Zfinal}. The importance of including such Jacobians to obtain the correct covariant measure has been stressed in \cite{Vassilevich:1994cz} for the case of QED.

\section{The off-shell effective action and gauge independence} \label{SemiClass}

It is well known that the off-shell effective action in quantum gravity generally depends on both the gauge and the field parameterisation (see e.g. \cite{buchbinder1992effective}).  
On the other hand since we observe which modes survive in the on-shell semiclassical approximation, and which modes cancel in the functional measure, we understand which fields are carrying physics, and 
which fields are only present due to us necessarily breaking diffeomorphism invariance. 

In order not to be forced to expand around a solution to the bare equations of motion we will now drop the condition \eq{onshell} and instead simply require that the background is an Einstein space,
\beq \label{Einstein}
R_{\mu\nu} = \frac{1}{d} g_{\mu\nu} R \,.
\eeq
As pointed out in the introduction this is crucial if we are to be able to extract the renormalisation of Newton's constant. 

\subsection{Gauge independence}
\label{Gindep}

Our aim is to preserve the gauge independence of the on-shell functional integral \eq{Zfinal} such that physically meaningful results may be derived off-shell.  
The problem that we face is that by working off-shell the cancellations between the measure and the $2d$ unphysical degrees of freedom is not guaranteed. For a generic parameterisation terms proportional to the equation of motion arise in the hessians and in the presence of such terms the cancellations observed in the previous section will not occur. In particular the hessians have the general form,
 \beq
 S^{(2)} = \tilde{S}^{(2)}+ X \left(R- \frac{2d}{d-2} \bar{\lambda}_b \right)\,,
 \eeq
 for any arbitrary parameterisation, where $X$ is a matrix in field space that depends on the parameterisation given by
 \beq
 X =  -\frac{d-2}{2d} \frac{\partial S^{(2)}}{\partial \bar{\lambda}_b} \,.
 \eeq
 which then implicitly defines $\tilde{S}^{(2)}$. While $X$ is dependent on the parameterisation $\tilde{S}^{(2)}$ is parameterisation independent.

At this point we make two observations, if $X \neq 0$ the gauge fixing dependence will remain and furthermore $S^{(2)}$ will possess negative eigenmodes depending on $\lambda_b$.
Thus, only by picking a parametrisation for which $X=0$ will the gauge fixing dependence vanish and the Gaussian integrals converge (assuming positivity of $S^{(2)}$ for the chosen background). By picking a parameterisation for which $X\neq 0$ the corresponding effective action will not formally exist for certain values of  $\bar{\lambda}_b$ where negative eigenvalues of $S^{(2)}$ are present, as well as being gauge dependent. We are therefore led to the conclusion that a parameterisation for which $X=0$ is the key to obtaining a physical meaningful gauge independent one-loop result. 
In turn this removes any dependence of the hessians on the cosmological constant.  \footnote{We note that the dependence on the cosmological constant can also be removed by use of specific gauge fixing conditions \cite{Percacci:2015wwa} whereby the conformal fluctuations $\bar{\phi}$ are constrained.}

To find such a parametrisation we first note that by using the exponential parameterisation \eq{expo} the equation of motion will only appear in the hessians for $\psi$ and $\phi$. In fact, if we revert to the the parameterisation in terms of $\bar{\psi}$ and $\bar{\phi}$, (see \eq{DBredef}) the equation of motion will only appear in the hessian for $\bar{\phi}$ due to \eq{detg}, this field being identified as the conformal factor of the metric. Then we may choose to parameterise this scalar degree of freedom in terms of a field $\sigma$ related to the conformal factor by
\beq \label{phitof}
e^{\frac{\bar{\phi}}{d}} = \left(1 + \frac{\sigma}{2}\right)^{\frac{2}{d}} = 1 + \frac{\sigma}{d} + \frac{2-d}{4 d^2} \sigma^2 + ...
\eeq
such that 
\beq \label{newpara}
\g_{\mu\nu} =  \left(1 + \frac{\sigma}{2}\right)^{\frac{2}{d}}   g_{\mu\lambda} [e^{\hat{h}}]^{\lambda}_\nu\,  = g_{\mu\nu} + \hat{h}_{\mu\nu} + g_{\mu\nu} \frac{\sigma}{d} + ... \,,
\eeq
where $\hat{h}_{\mu\nu}$ is a traceless fluctuation which may be decomposed as \eq{TT} with $\bar{\phi} = 0$.
It then follows that $X=0$ since the volume element is linear in the field and hence no term proportional to the cosmological constant can arise in the Hessian.

 It should be noted that the transformation between $\bar{\phi}$ and $\sigma$ is singular at the point $\sigma = -2$ and at this point $\gamma_{\mu\nu}$ vanishes. Thus the full non-perturbative treatment of the functional measure would require extra care involving the determinant of the field space metric to ensure that the measure is re-parameterisation invariant. In addition one should decide over which domain the functional integral should be performed and take care of possible singularities arising for a given parameterisation, such as at $\sigma = -2$.
Here we are concerned only with the semiclassical theory based on the expansion in small $\sigma$ to second order for which we can neglect the Jacobian which results from \eq{phitof}. One should also stress the choice \eq{phitof} is not unique. All that we require for $X$ to vanish is that $\sqrt{\g}$ has no term quadratic in the field
\beq \label{X0}
X=0    \iff  \left. \frac{\delta^2}{\delta h_{\mu\nu}(x) \delta h_{\rho \sigma}(y) } \int d^dx \sqrt{\gamma} \right|_{h_{\mu\nu}=0} = 0 \,.
\eeq 
 For any such choice the same semi-classical functional integral will be obtained. Additionally it may be natural to have all higher derivatives of the spacetime volume $V \equiv \int d^dx \sqrt{\g}$ to vanish as is the case for \eq{newpara}. In this case we have that all terms proportional to the equation of motion are absent in all higher vertex functions $S^{(n)}$ for $n\geq 2$.


Adopting \eq{newpara} we then perform the transformations \eq{DBredef} but now for $\sigma$ instead of $\bar{\phi}$,
\beq \label{DBredef2}
\sigma = \phi + \nabla^2 \bar{\psi}\,,  \,\,\,\,\,\,\,  \bar{\psi} = \psi + \frac{\rho}{(d-1-\rho) \nabla^2 + R} \phi \,.
\eeq
After this transformation the hessians then reduce to the form \eq{S2} and  \eq{S2phi} without use of the equations of motion. 
One may expand around an arbitrary Einstein background \eq{Einstein} and again obtain a functional integral given by \eq{Zfinal}, the upshot being that the scalar curvature $R$ is no longer fixed by the cosmological constant.

 The one-loop effective action $\Gamma$  can then be obtained from \eq{Zfinal} in the standard way by adding a source term and performing a Legendre transform. We then set $ \langle h_{\mu\nu} \rangle=0$ to obtain a functional of a single metric field $g_{\mu\nu}$.
 On an arbitrary Einstein space the gauge independent action $\Gamma$ then takes the simple form
 \beq \label{Gammaeff}
\Gamma[g_{\mu\nu}] - S[g_{\mu\nu}] =    \frac{1}{2} {\rm STr} [ \log S^{(2)}  ] =   \frac{1}{2} {\rm Tr}_{2T^2} [ \log \Delta_2  ]-\frac{1}{2} {\rm Tr}_{1T} [ \log \Delta_1  ] + \frac{1}{2} \log{\frac{d-2}{2 d^2} R} \,,
\eeq
 where $S$ is the total bare action appearing in the functional integral, with all determinants expressed in terms of auxiliary fields and ghosts, and ($ {\rm STr}$) {\rm Tr}  denotes the (super)-trace. The subscripts $2T^2$ and $1T$ denote that the traces are over transverse-traceless tensor modes and transverse vector modes, respectively, and the final term is the contribution of the constant mode of $\phi$. The simplicity of this effective action compared other gauge fixed actions is quite appealing, in particular it is independent of any scalar modes (apart from the constant mode $\phi_0$). Most importantly \eq{Gammaeff} has the property that all unphysical polarisations of the graviton, that cancel on-shell, also cancel off-shell. 
 
\subsection{Gauge invariance}
\label{Ginvar}

One may guess that the gauge independence which is observed when using the parameterisation \eq{phitof} is linked to an underlying gauge invariance.
In turn the lack of gauge independence for other parameterisations such as  \eq{linear} or \eq{expo} is due to the absence of gauge invariance.
In fact this can be observed by looking at the quadratic action  
\beq
S_{\rm quad} = \frac{1}{2} h \cdot S^{(2)}(g_{\mu\nu}) \cdot h \,
\eeq
for \eq{linear} or \eq{expo}, where we use a condensed notation. Apply a gauge transformation
\beq
h_{\mu\nu}  \to h_{\mu\nu} + \nabla_\mu \varepsilon_\nu + \nabla_\nu \varepsilon_\mu \,.
\eeq
One can then show that $S_{\rm quad}$ is only gauge invariant if the metric $g_{\mu\nu}$ is chosen to be on-shell.
This observation has been made by Deser and Henneaux  \cite{Deser:2006sq} for the linear parameterisation .
However if one instead takes the parameterisation \eq{newpara} we will be lead to a quadratic action of the form
\beq \label{quadraticvarphi}
S_{\rm quad} = \frac{1}{2} \varphi \cdot S^{(2)}(g_{\mu\nu}) \cdot \varphi \,
\eeq
with $\varphi = \{\sigma,\hat{h}_{\mu\nu}\}$. One may then show that this action is invariant under the linear gauge transformations
{\beq
\hat{h}_{\mu\nu}  \to \hat{h}_{\mu\nu} + \nabla_\mu \varepsilon_\nu + \nabla_\nu \varepsilon_\mu - \frac{2}{d} \nabla_\alpha \varepsilon^\alpha  \,,\,\,\,\,\,\,\,\,  \sigma \to \sigma + 2 \nabla_\alpha \varepsilon^\alpha \
\eeq
for an arbitrary Einstein space \eq{Einstein}. Thus the gauge independence of the one loop effective action \eq{Gammaeff} can be traced precisely to the gauge invariance of the underlying quadratic action \eq{quadraticvarphi} resulting from the parametrisation \eq{newpara}.

\subsection{de Donder gauge and the effective action for unconstrained fields}
\label{deDonder}

We have shown that a gauge independent one-loop effective action can be found at least for the gauges \eq{gf}. However, the understanding that this gauge independence is due to an underlying gauge invariance gives confidence that the effective action should be the same for other gauge choices outside the two parameter family \eq{gf}. To show this gauge independence we have utilised the transverse-traceless decomposition \eq{TT} which also leads to the Jacobians \eq{Js}. For consistency we now derive the effective action without use of the decomposition  \eq{TT} or the corresponding decomposition for the ghosts \eq{ghdecom} while working explicitly in the de Donder gauge for which we have $\rho =\frac{d}{2} -1$ and $\alpha =1$ in \eq{gf}. This gauge choice has been studied many times for the linear parameterisation and also recently for the exponential parameterisation \cite{Nink:2014yya}. Here we will use the parameterisation \eq{newpara}.
The hessians for the metric fluctuations then take the form
\beq \label{S2sigma}
S^{(2)}_{\sigma \sigma} =  \frac{d-2}{4d} \left|-\nabla^2 - \frac{2}{d} R \right| \, \equiv  \frac{d-2}{4d} |\Delta_s|  ,
\eeq   
\beq \label{S2hat}
S^{(2)}_{\hat{h} \hat{h}} = \frac{1}{2} \Delta_2 
\eeq
where we have Wick rotated  $\sigma \to i \sigma$ to give  $S^{(2)}_{\sigma \sigma}$ the positive sign for all modes apart from the constant mode. The corresponding ghost operator is given by
\beq \label{deDongh}
\det \mathcal{Q} = {\det}_1 \Delta_1 \,.
\eeq
Here we note that the neither the metric fluctuations nor the ghosts have any differential constraints. One then obtains the following 
one-loop effective action
\beq \label{GammadeDon}
\Gamma[g_{\mu\nu}] - S[g_{\mu\nu}]  =  \frac{1}{2} \Tr_{T2} \log \Delta_2 + \frac{1}{2} \Tr_0 \log |\Delta_s| - \Tr_1 \log \Delta_1 
\eeq 
where the first trace $\Tr_{2T}$ is over traceless tensor fluctuations, the second trace  $\Tr_{0}$  is over scalar fluctuations and the last trace $\Tr_1$ is over vector modes.
This effective action has a more familiar form of metric fluctuations minus ghosts fluctuations. 
Although its form is different from that of \eq{Gammaeff} one can use the relation between traces of differentially constrained fields and of non-differentially constrained fields to show that \eq{GammadeDon} and \eq{Gammaeff} are equivalent for $d>2$. Explicitly one has that the trace over transverse traceless modes is given by a trace over  traceless modes minus a trace over vector modes 
\beq \label{TT2}
\Tr_{2T^2} f (\Delta_2) =   \Tr_{2T} f(\Delta_2) - \Tr_1 f( \Delta_1)
\eeq
and that a trace over transverse vector modes is given by a trace over vectors minus a trace over scalars
\beq \label{T1}
 \Tr_{1T} f(\Delta_1) =   \Tr_1f( \Delta_1)  - \Tr_0' f( \Delta_s)  \,.
\eeq
Here the scalar trace has the constant mode removed. These relations \cite{Benedetti:2009gn, Lauscher:2001ya} hold on a generic Einstein space where we assume that there are no (conformal) killing vectors. One can then use \eq{TT2} and \eq{T1} to show that \eq{GammadeDon} is indeed equal to \eq{Gammaeff}.
To bring the effective action into it's simplest form we can use the relation between traceless tensor fluctuations and an unconstrained symmetric  tensor
\beq \label{T2}
\Tr_{2}f(\Delta_2) = \Tr_{2T} f(\Delta_2) + \Tr_0 f( |\Delta_s|)
\eeq
where $\Tr_{2}$ is the trace over symmetric two tensor modes with eigenvalue of constant mode sign flipped. Finally we get the expression
\beq \label{GammaUncons}
\Gamma[g_{\mu\nu}] - S[g_{\mu\nu}] = \frac{1}{2} \Tr_{2} \log \Delta_2  - \Tr_{1} \log \Delta_1
\eeq
which has the intuitive form of a trace over symmetric tensor fluctuations minus the ghost fluctuations with a relative factor of two in front of the ghosts without any further constraints for either field.  An alternative route to \eq{GammaUncons} can be found by introducing a fluctuation $h_{\mu\nu}$ at the beginning as 
\beq
\hat{h}_{\mu\nu} = h_{\mu\nu} - \frac{1}{d} h g_{\mu\nu} \,,\,\,\,\,\,   \sigma =  h^{\mu}\,_{\mu}  \,,
\eeq
such that we have just a single symmetric fluctuation field $h_{\mu\nu}$.
Then one finds the hessian in de Donder gauge is given by
\beq \label{S2hat}
S^{(2)}_{hh} = \frac{1}{2} \mathcal{C} \cdot \Delta_2 
\eeq
where the contribution to the effective action from the local operator $\mathcal{C}^{\mu\nu\lambda \rho} = \frac{1}{2} (g^{\mu\lambda}g^{\mu\rho} + g^{\mu\rho}g^{\mu\nu} - g^{\mu\nu} g^{\rho \lambda}) $ can be absorbed into the functional measure.

The effective actions \eq{GammaUncons}, \eq{GammadeDon} and \eq{Gammaeff} are all equivalent on a generic Einstein space. Care must be taken to treat Killing vectors and conformal Killing vectors if a background metric is chosen where they are present. In this paper we will side set this subtlety by assuming that no such modes are present.

\section{Interlude: Parameterisation independence of the Newtonian potential}\label{Low_energy}
The main aim of this paper is to compute beta functions which are independent of the gauge fixing employed. This we take as a prerequisite for the running couplings to be considered as physically meaningful. To find the beta function for Newton's constant we will look at the local divergencies of the effective action, however it is the non-local terms in the effective action which can be related to low scale observables in quantum gravity \cite{Donoghue:1993eb, Donoghue:1994dn}. When computing an observable the dependence on both the gauge and parameterisation should fall out of the result even though a dependence will generally be observed at intermediate stages of the calculation.

To check that the observables are indeed parameterisation independent we will calculate the corrections to the Newtonian potential felt by a (classical) test particle via corrections to its geodesic equation \cite{Dalvit:1997yc}, generalising the result to an arbitrary parameterisation. These corrections come from two sources, namely the quantum correction to the Einstein equations and the quantum corrections to the geodesic equation itself. Only when both of these are taken into account does the dependence on the gauge and parameterisation cancel out. Other approaches to defining a physical Newtonian potential have been considered such as \cite{BjerrumBohr:2002kt} based on scattering potential, however the explicit gauge independence has not been shown.

Here we consider a classical source of mass $M$ and a test particle of mass $m$ coupled to gravity with a bare action of the form
\beq \label{Sparticles}
S = \int d^4x \sqrt{-\gamma} \frac{R}{16 \pi G_b} - M\int  \sqrt{ - \gamma_{\mu\nu} dy^\mu dy^\nu} - m\int  \sqrt{ - \gamma_{\mu\nu} dz^\mu dz^\nu}
\eeq
where it is assumed that $m \ll M$.  The classical field Einstein equations therefore involve a non-zero energy momentum tensor $T^{\mu\nu} = T^{\mu\nu}_M +T^{\mu\nu}_m$.
 After computing the one loop effective action quantum corrections to the Newtonian potential $\mathcal{V}$ are found via the equation of motion for $z$ in the non-relativistic limit
\beq \label{defPot}
 - \vec{\nabla} \mathcal{V} = \frac{d^2 \vec{z}}{dt^2}
\eeq
where the particle of mass $M$ is taken to be a stationary source.

In \cite{Dalvit:1997yc} the potential $\mathcal{V}$ was calculated for a one parameter family of gauges for which $\rho=1$ and $\alpha$ is the free parameter. There the linear parameterisation was used whereas here we wish to analysis the parameterisation dependence for a fixed gauge. To this end we consider the four parameter family of metrics given 
\beq \label{genpara}
\gamma_{\mu\nu} = g_{\mu\nu} + h_{\mu\nu} + \beta_1\, h_{\mu}\,^\lambda h_{\lambda \nu} + \beta_2\, h_{\mu\nu} h^\lambda\,_\lambda + \beta_3 \, g_{\mu\nu} h^{\lambda \rho} h_{\lambda \rho} + \beta_4 \, g_{\mu\nu} h^{\lambda}\,_\lambda h^\rho\,_\rho
\eeq 
while fixing the gauge to $\rho=1$ and $\alpha =1$. The parameterisation \eq{newpara} corresponds to the choice $\beta_1 = 1$, $\beta_2 = -\frac{1}{d}$, $\beta_3 = 0$, $\beta_4 = \frac{2-d}{2 d^2}$. Here we repeat the work of \cite{Dalvit:1997yc} recalling only the main results and stating the differences arising from taking $\beta_i \neq 0$. 

To find the corrections to the Newtonian potential we need to calculated the non-local terms in the effective action. First the quantum corrections due to the heavy particle and gravity can be found neglecting the term proportional to $m$.
With $\beta_i = 0$ these terms take the form $\Gamma^{\rm nl}(\beta_i =0) = \Gamma^{\rm nl}_G(\beta_i =0) + \Gamma^{\rm nl}_M(\beta_i =0)$  \cite{Dalvit:1997yc}
\bea
\Gamma^{\rm nl}_G(\beta_i =0) &=& - \frac{1}{96 \pi^2} \int d^4x \sqrt{-g} \left[ \frac{21}{10} R_{\mu\nu} \log(-\Box) R^{\mu\nu} + \frac{1}{20} R \log(-\Box) R \right] \,, \\
\Gamma^{\rm nl}_M(\beta_i =0) &=& -\frac{1}{64 \pi^2} \int d^4x \sqrt{-g} \left[ M_{\mu\nu\rho \sigma} \log(-\Box) M^{\rho \sigma \mu \nu}+ 2 M_{\mu\nu \rho \sigma} \log(-\Box) \left( P^{\rho\sigma \mu \nu} + \frac{1}{6} R g^{\rho \left(\mu \right.} g^{\left. \nu\right) \sigma} \right) \right] \,. \nonumber
\eea
where the tensors $P$ and $M$ are given by
\bea
P^{\mu\nu}_{\lambda \sigma} = 2 R_\lambda\,^{(\mu}\,_{\sigma}\,^{\nu)} + 2 \delta^{(\mu}_{(\lambda} R^{\nu)}_{\sigma)} - g^{\mu\nu} R_{\lambda \sigma} - g_{\lambda \sigma} R^{\mu\nu} - R \delta^\mu_{(\lambda} \delta^{\nu}_{\sigma)} + \frac{1}{2} g^{\mu\nu} g_{\lambda \sigma} R \\
M^{\mu\nu}_{\lambda \sigma} = M^2  4 \pi G_N \int d\tau \delta^4(x- y(\tau))[g^{\mu\nu} \dot{y}_\lambda \dot{y}_\sigma + 2 \dot{y}^\mu \dot{y}^\nu \dot{y}_\lambda  \dot{y}_\sigma] 
\eea

Here we have calculated the non-local terms for general parameterisation finding  $\Gamma^{\rm nl} = \Gamma^{\rm nl}(\beta_i =0) + \Gamma^{\rm nl}_\beta$ 
\beq
\Gamma^{\rm nl}_\beta = \int d^4 x \sqrt{-g} [A(\beta_i) R_{\mu\nu} \log(-\Box) \mathcal{E}^{\mu\nu} + B(\beta_i) R g_{\mu\nu} \log(-\Box)  \mathcal{E}^{\mu\nu} ]
\eeq 
with
\beq
A = - \frac{\beta_1 G_N}{32 \pi}\,,\,\,\,\,\,\,\,\,\, B = -\frac{G_N (439 \beta_1 +58 \beta_2+1768 \beta_3+232 \beta_4 )}{384 \pi } \,.
\eeq
As expected the effective action for different parameterisations only differs by terms proportional to the field equations $\mathcal{E}^{\mu\nu} = -\frac{1}{16 \pi G_N} (R^{\mu\nu} - \frac1 2 R g^{\mu\nu}) + \frac1 2 T^{\mu\nu}_M$ (here we neglect terms quadratic in the equation of motion). In fact the term $\Gamma^{\rm nl}_\beta$ has the same form as terms which arise from choosing a different gauge $\alpha \neq 1$ where instead $A$ and $B$ would depend on the gauge parameter.

Solving the effective Einstein equations for a static source $ \dot{y}^\mu = (1,0,0,0)$ in the Newtonian limit one finds that the $00$-component of the perturbation $\bar{h}_{\mu\nu}$ is given by
\beq \label{h00}
\bar{h}_{00} = \frac{2 G M}{r} \left[ 1 + \frac{43 G_N}{30 \pi r^2} - \frac{5G}{12 \pi r^2} + \frac{A(\beta_i) - 2 B(\beta_i)}{r^2}  \right]
\eeq
which depends on the parameterisation through $A$ and $B$. The second and third terms in \eq{h00} are the quantum corrections for $\beta_i =0$ due to gravitons and the heavy particle respectively.  The crucial step is to now find the quantum corrections to the geodesic equation for the test particle of mass $m$. These arise since the  particle couples to the quantum metric itself and thus the effective action will also include an extra contribution $\Gamma_m^{\rm nl}$ proportional to $m$. This term is given by
\bea
\Gamma_m^{\rm nl} &=& \int d^4x \sqrt{-g} \left[ -\frac{1}{32 \pi^2} m_{\mu\nu\rho\sigma} \log(-\Box) M^{\rho\sigma \mu\nu} - \frac{1}{32 \pi^2} m_{\mu\nu\rho\sigma} \log(-\Box) \left(  P^{\rho\sigma \mu\nu} + \frac{1}{6} R g^{\rho \left(\mu \right.} g^{\left. \nu\right) \sigma} \right) \right. \nonumber \\
&+& \left.  \frac{A(\beta_i)}{2} R_{\mu\nu} \log(-\Box) T^{\mu\nu}_m + \frac{B(\beta_i)}{2} R g_{\mu\nu} \log(-\Box) T^{\mu\nu}_m \right]
\eea
with $m_{\mu\nu\rho\sigma}$ equal to  $M_{\mu\nu\rho\sigma}$ with the replacement $M \to m$ and $y^\mu \to z^\nu$

In the non-relativistic limit the quantum corrected geodesic equation is given by
\beq
\frac{d^2 \vec{z}}{dt^2} -  \frac{1}{2} \vec{\nabla} h_{00} = \frac{1}{m} \frac{\delta \Gamma^{\rm nl}_m}{ \delta \vec{z}}
\eeq   
where both $ \Gamma^{\rm nl}_m$ and $h_{00}$ depend on the parameterisation. However taking into account both of these contributions one finds that $\frac{d^2 \vec{z}}{dt^2}$  is independent of the parameterisation. Explicitly one finds that the potential \eq{defPot} is given by
\beq \label{potential}
\mathcal{V}(r) = - \frac{G_N M}{r} \left[ 1+  \frac{43 G_N }{30 \pi r^2} - \frac{5 G_N }{12 \pi r^2} + \frac{7 G_N }{12 \pi r^2} \right]
\eeq
where the effect of coupling the test particle is to cancel the parameterisation dependence and contribute the last term. 
Thus we observe that, unlike the beta functions, the Newtonian potential is independent of the parameterisation of the quantum fields.

\section{Semi-classical flow equation}  \label{flow}
We now return to the main topic of this paper namely obtaining a gauge independent beta function for Newton's constant.
To this end we will use the results of sections~\ref{Measure} and \ref{SemiClass} utilising the parameterisation \eq{newpara}. The starting point is the functional integral \eq{Zfinal} on an arbitrary Einstein background with the constant mode $\phi_0$ removed (in $d>2$ this mode will not renormalise the Newton's constant but should modify curvature squared corrections in $d=4$).
Here we shall use the modern formalism of the non-perturbative renormalisation group \cite{Wetterich:1992yh,Morris:1993qb, Reuter:1996cp}, even though it is not technically needed since we are working at one-loop. This approach can then be compared to the old fashioned renormalisation scheme of simply evaluating \eq{Gammaeff} and regulating the one loop integrals.  To this end we add a regulator term, 
\beq
\Delta S_k[\varphi] = \frac{1}{2} \varphi \cdot \mathcal{R}_k \cdot \varphi\,,
\eeq
to the bare action. Here $\mathcal{R}_k$ is an infra red regulator which suppresses low momentum modes $p^2 < k^2$ while vanishing for high momentum modes $p^2 \gg k^2$.
The scale $k$ plays the role of $\mu$ as the scale which defines the beta functions. 
In the case where we use \eq{Gammaeff} the fields  $\varphi =\{ h^\upvdash, \zeta, \bar{c}, c \}$ are all fields appearing in the bare action after exponentiating the determinants in functional measure. Alternatively we can use the effective actions \eq{GammadeDon} or  \eq{GammaUncons}  for the unconstrained fields in which case we have $\varphi =\{ \hat{h}_{\mu\nu}, \sigma , \bar{C}_\mu , C_\mu\}$ or  $\varphi =\{ h_{\mu\nu}, \bar{C}_\mu , C_\mu\}$. In any case $\varphi$ can be thought of as a vector in field space with components $\varphi_a$ and $\mathcal{R}_k$ as a matrix with components $\mathcal{R}_{ab}$.

For the covariant momentum $p^2$ we take the eigenvalues of the differential operators of the operators $\Delta_n$ for $n= \{1,2, s\}$ (given in equations \eq{Deltas}, \eq{Delta2} and \eq{S2sigma}). The regulator then takes a form such that
\beq \label{RegChoice}
S^{(2)}(\Delta_n) + \mathcal{R}_k(\Delta_n) = S^{(2)}(\Delta_n \to P_n) \,,\,\,\,\,\,\,\,\, P_n =  \Delta_n + k^2 C(\Delta_n/k^2)\,,
\eeq
where $C(z)$ is a dimensionless scalar cutoff function which vanishes for $z\gg1$ but stays finite and non-zero for $z \to 0$. This constitutes a type-II cutoff, in the nomenclature of \cite{Codello:2008vh}, where also the potential terms of $\Delta_{n}$ are included in the regulator. Another choice, termed type I, is to replace $\Delta_n$ by $-\nabla^2$ in \eq{RegChoice}, however we expect a non-trivial cancellation to occur between the traces of the transverse traceless fluctuations and the Jacobian $J_1$ such that only $(d-3) d/2$ propagating degrees of freedom remain. Since, without the regulator $\mathcal{R}_k$, these cancellations must be between traces depending only on the operators $\Delta_{n}$, that appear in \eq{Gammaeff}, it seems natural that the regulator should also depend solely on these operators as well. In addition it has been shown \cite{Dona:2012am} that only type II regulators for fermions correctly regulate modes for the Dirac operator in curved spacetime. Finally we point out that it is the properties of the Lichnerowicz Laplacian $\Delta_2$ that determine the stability of classical solutions to the Einstein equations and hence one should expect that this continues to be the case in the semiclassical theory.

After inserting the regulator into the path integral the flowing effective action $\Gamma_k$ is defined by,
\beq \label{IntGamma}
e^{-\Gamma_k[\bar{\varphi}]} = \int \mathcal{D} \varphi \, e^{-S[\varphi] -\Delta S_k[\varphi - \bar\varphi]   + (\varphi - \bar{\varphi}) \cdot \frac{\delta \Gamma_k }{\delta \bar\varphi} }\,,
\eeq
where $\bar{\varphi} = \langle \varphi \rangle$. One observes that if $\mathcal{R}_k$ diverges for $k^2 \gg p^2$ the regulator term becomes a Gaussian peaked around $\varphi = \bar{\varphi}$, whereas for $k^2 \to 0$ the regulator must vanish and \eq{IntGamma} assumes the form of the effective action. It follows that $\Gamma_k$ interpolates between the bare action $S$ for large $k$ and the effective $\Gamma$ in the limit $k\to 0$.  In the semi-classical approximation the flowing effective action takes the form,
\beq \label{oneloopaction}
\Gamma_k - S [\bar\varphi] = \frac{1}{2} {\rm STr} \log \left( S^{(2)} + \mathcal{R}_k \right)\,.
\eeq
Due to the relationships between the traces \eq{TT2}, \eq{T1} and \eq{T2} and the choice of the regulator \eq{RegChoice} the equivalence between the effective actions $\Gamma$ also holds for the $k$ dependent actions
\bea \label{oneloopaction}
\Gamma_k - S [\bar\varphi]& =&   \frac{1}{2} {\rm Tr}_{2 T^2} \left[ \log \left(\Delta_2 +   \mathcal{R}_k(\Delta_2) \right) \right]-\frac{1}{2} {\rm Tr}_{1 T} \left[ \log\left( \Delta_1 +  \mathcal{R}_k(\Delta_1) \right)  \right] \nonumber \\
&=& \frac{1}{2} {\rm Tr}_{2} \left[ \log \left(\Delta_2 +   \mathcal{R}_k(\Delta_2) \right) \right]- {\rm Tr}_{1} \left[ \log\left( \Delta_1 +  \mathcal{R}_k(\Delta_1) \right)  \right]\,,
\eea 
where we have absorbed irrelevant constant factors of $16 \pi G_b$ into the fields. 
 Setting $\mathcal{R}_k =0$ we recover \eq{Gammaeff}.
 Taking a derivative with respect to the Wilsionian RG time  $t = \log k/k_0$ to obtain the following gauge independent flow equation, 
\bea \label{oneloopflow}
\partial_t \Gamma_k &=& \frac{1}{2} \Tr_{2 T^2} \frac{\partial_t \mathcal{R}_k(\Delta_2)}{ \Delta_2 + \mathcal{R}_k(\Delta_2)} -\frac{1}{2} \Tr_{1 T} \frac{\partial_t \mathcal{R}_k(\Delta_1)}{\Delta_1 + \mathcal{R}_k(\Delta_1)} \nonumber \\
&=& \frac{1}{2} \Tr_{2} \frac{\partial_t \mathcal{R}_k(\Delta_2)}{ \Delta_2 + \mathcal{R}_k(\Delta_2)} - \Tr_{1} \frac{\partial_t \mathcal{R}_k(\Delta_1)}{\Delta_1 + \mathcal{R}_k(\Delta_1)} 
\eea 
 Note that the flow equation \eq{oneloopflow}, unlike \eq{Gammaeff}, is both UV and IR finite and therefore does not necessitate the introduction of an explicit UV cutoff. Instead the bare action enters the flow equation as a boundary condition. Including the constant mode $\phi_0$ we give an extra term in he first line of \eq{oneloopflow} arising from this single mode. In the second line of \eq{oneloopflow} we give the form of the flow equation for unconstrained fields.
 
 Here we consider the following ansatz the flowing action 
 \beq
\Gamma_k = \frac{1}{16 \pi G_k} \int d^d x \sqrt{\bar{g}}  \left( 2\bar{\lambda}_k -  R \right) \equiv S[g_{\mu\nu}] +   \int d^d x \sqrt{\bar{g}}  \left(\delta \lambda_k -  \delta \kappa_k R\right) \,,
\eeq
which takes the Einstein Hilbert form and allows us to extract the beta functions for the running couplings $G_k$ and $\bar{\lambda}_k$.
Equivalently we define the vacuum energy density $\lambda_k$ and the inverse Newtons couplings $\kappa_k$ by 
\beq
\lambda_k \equiv \frac{\bar{\lambda}_k}{8 \pi G_k} \equiv \frac{\bar{\lambda}_b}{8 \pi G_b} +  \delta \lambda_k\,,\,\,\,\,\,\,\,\,\,\,\,  \kappa_k \equiv \frac{1}{16 \pi G_k} \equiv \frac{1}{16 \pi G_b} + \delta \kappa_k\,.
\eeq
The UV scale $\Lambda$ can then be defined as the scale for which $\delta \lambda_{k=\Lambda}=0=\delta \kappa_{k= \Lambda}$. In turn these boundary definitions can be viewed as replacing the operational meaning of the `bare quantities' $G_b$ and $\bar{\lambda}_b$ entering the functional integral. Thus, the flow equation defines a renormalisation scheme whereby the UV boundary condition replaces the bare action and the IR boundary condition sets the renormalisation condition in terms of renormalised quantities obtained in the limit $k=0$. The advantage of this scheme is that it dispenses with with the formally divergent path integral and generalises beyond the perturbative regime \cite{Wetterich:1992yh,Morris:1993qb, Reuter:1996cp}.

We now put the flow equation together expressing all quantities in units of $k$ in order to find the autonomous system of beta functions. In particular we drop the index $k$ from the couplings when referring to the dimensionless quantities such that $G = k^{d-2} G_k$ and $\bar{\lambda} = k^{-2} \bar{\lambda}_k$. The dimensionless flow equation then takes the form
\beq \label{dimensionlessflow}
\int d^dx \sqrt{g}\left( \frac{\partial_t G - (d-2) G}{ 16 \pi G^2} R  + \partial_t \lambda  + d \lambda \right) = \Tr_2 [W(\Delta_2)]  -2 \, \Tr_1 [W(\Delta_1)] \,,
\eeq
where we define the function,
\beq
W(z) =  \frac{  C(z) -  z  C'(z)    }{ z +C(z)} \,.
\eeq
This gauge independent flow equation \eq{dimensionlessflow} is the main result of this section.

\section{Heat kernels and universality} \label{HeatKernels}
To obtain the beta functions one  computes the traces on the RHS of the flow equation \eq{dimensionlessflow} and compares the terms on each side of the equation to linear order in the curvature. In the same way one way evaluate \eq{Gammaeff} directly to determine the counter terms needed for the one loop effective action on an Einstein space.
To evaluate the traces in both cases we therefore exploit the early time heat kernel expansion for the operators $\Delta_1$ and $\Delta_2$ acting on the vectors and symmetric tensors respectively.
 For a general function $f(\Delta_n)$ one has the following expressions,
\beq
\Tr_n [f(\Delta_n)] = \frac{1}{(4 \pi)^{\frac{d}{2}}}\sum_{i=0}^{\infty} Q_{\frac{d}{2}-i}[f]  \int d^dx \sqrt{g} a_{2i,n}\,,
\eeq
where the functionals $Q_m$ for $m>0$  are given by
\beq
Q_m[f] = \frac{1}{\Gamma(m)}\int^\infty_0 dz z^{m-1} f(z)\,,
\eeq
and the coefficients $A_{2i,n}$ are the heat kernel coefficients for the operators $\Delta_n$ proportional to curvature monomials with mass dimension $2i$.
In order to find the beta functions for $G$ and $\Lambda$ we need only the heat kernels for $i = 0,1$. For $i = 0$ the heat kernel coefficients are equal to the dimension of the field space
\beq
a_{0,1} = d  \,,  \,\,\,\,\,\,\, a_{0,2} = \frac{d(d+1)}{2}  
\eeq
whereas for $i =1$ the heat kernel coefficients of a generic operator of the form
\beq \label{deltaE}
\Delta_n = -(\nabla^2 + E_n) \,,\,\,\,\,\,\,\, \rm{with} \,\,\,\,\, (E_2)^{\mu\nu}\,_{\alpha\beta} = 2 R^\mu\,_{\alpha}\,^\nu\,_\beta \,, \,\,\, (E_1)^\mu\,_{\alpha} =  R^\mu\,_\alpha
\eeq
are given by
\beq
a_{2,n} = \frac{1}{6} d_n R + \tr_n E_n  
\eeq
where the trace $\tr_n$ denotes the trace over the field space leading to $\tr_1 E_1  = R$ and $\tr_2 E_2 
 =  -R$
We then obtain
\bea \label{evaluatedtraces}
\Tr_2 [W(\Delta_2)]  - 2 \Tr [W(\Delta_1)]  &=& \frac{1}{(4 \pi)^{\frac{d}{2}}}  \left(  (a_{0,1} -  2 a_{0,2})   \, Q_{\frac{d}{2}}[W]   +  \left( \frac{1}{6} N_g + \tr_2 E_2 - 2 \tr_1 E_1 \right) Q_{\frac{d}{2}-1}[W]  \right)  \\
&=& \frac{1}{(4 \pi)^{\frac{d}{2}}} \int d^dx \sqrt{g}  \left(  N_g \, Q_{\frac{d}{2}}[W]   + \frac{1}{6} ( N_g - 18) \, Q_{\frac{d}{2}-1}[W]  R \right) + ...  
\eea
where $N_g= \frac{1}{2} d (d-3)$ is the number of propagating degrees of freedom in $d$-dimensional quantum gravity.
One sees here the universal origin of the factor  $N_g - 18$ coming just from the heat kernel coefficient.
We note that one could of have simply renormalised via the introduction of counter terms without the introduction of the IR regulator. That is we could evaluate \eq{Gammaeff} directly to obtain
\bea \label{evaluatedtracesGamma}
\Gamma 
&=&S+ \frac{1}{2(4 \pi)^{\frac{d}{2}}} \int d^dx \sqrt{g}  \left(  N_g \, Q_{\frac{d}{2}}[\log]   + \frac{1}{6} ( N_g - 18) \, Q_{\frac{d}{2}-1}[\log]  R \right) + ...  
\eea
Then one simply observes that the counter term for the $\sqrt{\g} R(\g)$ term in \eq{action} would be proportional to  $N_g - 18$ and the volume term is proportional to $N_g$. Thus we see, from this point of view, why it is important to make the regulator a function of the differential operators appearing in the hessians themselves. Any other choice, such as $\mathcal{R}_k=\mathcal{R}_k(-\nabla^2)$, would lead to non-universal heat kernel coefficients that do not appear in the one-loop effective action. One should also stress that the origin of   $-11 + \frac{2}{3} N_f$ can also be traced to the heat kernel coefficient for the corresponding tensor structure in Yang-Mills \cite{Reuter1994181}. In addition one my also go to the next order in the curvature expansion where in $d=4$ the result is universal since $Q_0[W] = W(0) =1$. Here we find explicitly the curvature squared term
\beq
\Tr_{2} [W(\Delta_2)]  - 2 \Tr_{1} [W(\Delta_1)]  = ... +  \frac{1}{(4 \pi)^2} \int d^dx \sqrt{g}  \left(  \frac{53}{45} R^{\mu\nu\rho\sigma} R_{\mu\nu\rho\sigma} -\frac{29}{40} R^2 \right) + ...
\eeq
reproducing the  one-loop counter term found in \cite{Christensen:1979iy} but with the cosmological constant replaced by $\bar{\lambda}_b \to R/4$. 
Here we note that the inclusion of the constant mode contribution (which does not contribute the the renormalisation of Newton's constant) is need to correctly evaluate these terms.  

We note the general pattern that universal factors appearing in beta functions and counter terms throughout quantum field theory can be found by use of heat kernel techniques. These factors can be found for dimensionful couplings, such as Newton's constant, as well as dimensionless ones. The difference is that the renormalisation of dimensionless couplings do not come additionally with non-universal factors. However the sign of non-universal factors appearing in the renormalisation of dimensionful should be universal and gauge independent if they are to be physically meaningful. The fact that the universal factor found here only depends on the dimension through the number of polarisations would seem to support the view that this is indeed the case.

It should be noted, however, that the unique factors $N_g - 18$ and $N_g$ could be modified if one was to regulate the auxiliary fields fields differently than the transverse traceless fields. Thats is if we choose different regulators $C_1(z)$ and $C_2(z)$ for the separate traces in \eq{dimensionlessflow} we would end up with different functions $W(z)$. Making such a choice would mean that the factors $N_g$ and  $N_g - 18$ of \eq{evaluatedtraces} no longer be forthcoming. Equally if we choose to regulate with a type I cut off such that $\mathcal{R}_k = \mathcal{R}_k(-\nabla^2)$ rather than the choice \eq{RegChoice} these factors would again not appear in \eq{evaluatedtraces}. One must conclude then that if the factor $N_g - 18$ is truly universal and physical only a sub set of all possible regulator choices leads to physical results. What we suspect is that regulator schemes must break diffeomorphism invariance sufficiently mildly such that the cancellations of unphysical polarisations continues to be present.

To back up our point of view imagine we pick $C_1(z) \neq C_2(z)$, then the term in \eq{evaluatedtraces} proportional to the space time volume would not necessarily be proportional to $d-3$. This being the case the vacuum energy would appear to be renormalised in $d=3$ even though the evaluation of the unregulated effective action \eq{evaluatedtracesGamma} would lead to no divergencies proportional to the volume. A similar argument can be made for the factor $N_g - 18$ in the critical dimension $d=d_c$ for which $N_g(d_c) = 18$. Thus only by picking the regulator scheme as we have do we avoid such inconsistencies.

\section{Beta function for Newton's constant} \label{betaGsec}

It is now straightforward to find the beta function $\beta_G = \partial_t G$ for Newton's constant by inserting \eq{evaluatedtraces} into \eq{dimensionlessflow}. For general cutoff function $C(z)$ it reads
\beq \label{betaG}
\beta_G = (d-2) G +  \frac{2}{3} \frac{N_g -18}{ (4 \pi)^{\frac{d}{2}-1} \Gamma(\frac{d-2}{2})} \, \, \mathcal{I}_{d/2 -1}[C] \, G^2
\eeq
where the regulator dependent functional $\mathcal{I}_n[C]$ is given by the integral
\beq \label{I}
 \mathcal{I}_n[C_k] \equiv \int_0^\infty dz  z^{n-1} W(z)  = \int_0^\infty dz z^{d/2 -2}  \frac{ C(z) -  z  C'(z)    }{ z +C(z)} \,.
\eeq
The first term in \eq{betaG} is the classical scaling arising from the dimensionful nature of $G_k$. The second term is a quantum correction proportional to the momentum integral   $\mathcal{I}_{d/2 -1}[C]$ indicating that this equation is one loop.

\subsection{ $d= 2+ \epsilon$ and  $d=2$ dimensions, a tale of three beta functions}

Before continuing to the general $d$ case we take a digression into the behaviour of the beta function near two dimensions and try to understand the subtleties of the limit $d \to 2$.
First we note that in the limit $d \to 2$ we have
\beq \label{Idto2}
 \left. \mathcal{I}_{d/2 -1}[C]\right|_{d \to 2} = \frac{2}{d -2} + {\rm finite\,terms}
\eeq
where the singular term is independent of the regulator function $C$. However the coefficient of $G^2$ has a finite universal limit for $d \to 2$ owing to the presence of the gamma function. This allows one to extract a universal beta function in $d = 2 + \epsilon$ dimensions and in the case $d=2$.

\subsubsection{$d=2 + \epsilon$ dimensions}

For $d = 2 + \epsilon$ we find the universal beta function 
\beq \label{betaep}
\beta_{G}  = \epsilon \, G - \frac{38}{3} G^2 \,.
\eeq 
We note that this is in agreement with previous studies \cite{Tsao:1977tj, Jack:1990ey, Kawai:1989yh} using the linear parameterisation \eq{linear} but differs from the result obtained with the exponential parameterisation \eq{expo} \cite{Kawai:1992fz,Nink:2014yya} with $\rho = d/2 -1$ and  $\bar{\lambda}_b = 0$ which gives
\beq  \label{betaepex}
\beta_{G}  = \epsilon \, G - \frac{50}{3} G^2 \,.
\eeq
The reason for this is that by working off-shell and setting $\bar{\lambda}_b = 0$ the exponential parameterisation gives a hessian for the scalar mode $\phi$ becomes $S^{(2)}_{\phi\phi} \propto (d-2) \Delta$ and thus the cancellation between these modes and $J_0$ does not occur. Instead the hessian for the conformal modes mimics the induced action 
\beq \label{Sanom}
S_{\rm anom} \propto \int d^2x \sqrt{g} ( \frac{\bar{\phi}}{4} R + \frac{1}{16} \bar{\phi} \Delta \bar{\phi})
\eeq
of the conformal anomaly in $d=2$ despite being of order $\epsilon$ \cite{Kawai:1992fz}.  In the equations leading to \eq{betaep} such fluctuations are not present since all conformal fluctuations are cancelled by the functional measure.
The result \eq{betaepex} can then be understood as arising from the particular way the limit $d \to 2$ is taken where the fact that the hessian for $\bar{\phi}$ vanishes is not taken into account. The limit then paradoxically reproduces the expected result in $d=2$ in the presence of the conformal anomaly even though the induced action \eq{Sanom} for has not been accounted for.

{ In order to help clarify the situation we briefly recall the derivations of \eq{betaep} given in \cite{Kawai:1989yh} and generalise it to an arbitrary parameterisation. The derivation differs in two respects from the methods used in this paper. Firstly dimensional regularisation is used such that only the logarithmic divergences are retained and secondly the freedom to redefine the background metric is exploited in order to define a physical beta function.   Thus first one obtains the one loop equation 
\beq
\partial_t \Gamma[g_{\mu\nu}] = \int d^4x \sqrt{g} \mu^{\epsilon}  \left[   - \frac{1}{16 \pi} \cdot   \frac{38}{3}  R  + \bar{b} \left(R - \frac{2d}{ \epsilon} \bar{\lambda} \right) \right]
\eeq
  where $\bar{b}= \bar{b}(\alpha, \rho,\beta_i)$ is a number which depends on the gauge \eq{gf} and parameterisation \eq{genpara} and $\mu$ is the renormalisation scale. For the special case \eq{newpara} one has $\bar{b}=0$. To remove the unphysical dependencies in the case $\bar{b} \neq0$ the background metric is redefined $g_{\mu\nu} \to g_{\mu\nu}' = Z_g g_{\mu\nu} $ in such a way that the cosmological constant with respect to  $g_{\mu\nu}'$ is not renormalised. Explicitly one has 
\beq
Z_g^{-1} \partial_t Z_g =  8 \pi G \mu^\epsilon \epsilon \,   \bar{b}
\eeq  
  leading to  
\beq
\partial_t \Gamma[g_{\mu\nu}'] = \mu^\epsilon \int d^4x \sqrt{g'} \left[    - \frac{1}{16 \pi}  \cdot \frac{38}{3} R'  \right]
\eeq
 reproducing \eq{betaep} independently of both the gauge and the parameterisation. At a deeper level one finds that the renormalisation of the dimensionless product $G \lambda^{\frac{\epsilon}{ \epsilon + 2}  }  $ is independent of $Z_g$. Indeed one may define the beta function via the relation between the bare and renormalised dimensionless products in units $\mu =1$
\beq \label{product}
G_b^d \lambda_b^{d-2} = G^d \lambda^{d-2} \left( 1 + \frac{38}{3}   \frac{d}{d-2} G \right)
\eeq
which is independent of $\bar{b}$ and hence both the gauge and parameterisation.
The beta function is then found by asking how $G$ should be changed if $\lambda \to \lambda + \delta \lambda$ while keeping $G_b^d \lambda_b^{d-2}$ fixed according to the definition \cite{Kawai:1989yh} 
\beq \label{physicalbeta}
\delta G = - \frac{1}{d} \frac{\delta \lambda}{\lambda} \beta_G \, .
\eeq
This beta function is again given by \eq{betaep}. Thus we observe that when considering the physical renormalisation there is no dependence on the parameterisation. The advantage of using a parameterisation or gauge for which $\bar{b}=0$ is that the na{\"i}ve beta function automatically agrees with the physical one. One should note that different physical beta functions may be defined once other couplings or masses are introduced \cite{Kawai:1989yh}.

\subsubsection{$d=2$ dimensions}

 One should note that the beta function \eq{betaep} is valid for $d \neq 2$ and that for the strictly $d=2$ the hessian for the modes $\phi$ vanishes reflecting  conformal invariance. To work explicitly in two dimensions we note that $\int d^2x \sqrt{\g} R$ is a topological invariant however the inclusion of both gauge fixing and ghost terms means that the path integral is non-trivial. Furthermore since there are no transverse-traceless fluctuations in two dimensions the use of the transverse-traceless decomposition is problematic. As such we will use de Donder gauge (i.e. for $\alpha=1$ and $\rho =0$) such that the hessians and ghosts determinants are given  \eq{S2sigma}, \eq{S2hat} and \eq{deDongh}.

 We observe then that there are both traceless metric and ghost contributions while the hessian of the conformal fluctuation vanishes. Since the second variation of the Einstein-Hilbert action itself vanishes in $d=2$ it is the case that \eq{S2hat} is the second variation of the gauge fixing action alone even though in higher dimensions it results from a cancellation of gauge fixing terms with terms in the second variation of $S_{\rm EH}=-\int d^2x \sqrt{\g} R$. To make sense of this we can observe that an explicit computation of $S^{(2)}_{\rm EH}=0$ for $d=2$ gives a relation which implies that the second variation of the gauge fixing action is
\beq
16 \pi G_b \, \hat{h} \cdot S^{(2)}_{{\rm gf}, \hat{h} \hat{h}} \cdot \hat{h} = -\int d^2x \sqrt{g} \,  \hat{h}^{\mu\nu} \nabla_\mu \nabla_\lambda \hat{h}^\lambda_\mu =  \frac{1}{2} \int d^2x \sqrt{g} \, \hat{h}^{\mu\nu} \Delta_2 \hat{h}_{\mu\nu} \,,
\eeq
leading to \eq{S2hat}. As a crosscheck the second equality above can be confirmed by writing $\hat{h}_{\mu\nu} = \nabla_\mu v_\nu + \nabla_\nu v_\mu - g_{\mu\nu} \nabla_\alpha v^\alpha$ which is a complete decomposition of a traceless symmetric two tensor in two dimensions.

If we additionally include $D$ scalar fields these can be interpreted as the $D$ spacetime dimensions in string theory. That is we include an action of the form
\beq
S_{\rm string} = \left. \int d^dx\, \sqrt{\g} G_{IJ} \partial_\mu X^I \partial^\mu X^J \right|_{d\to 2}\,,
\eeq
with $I= 1,..,D$ where we assume $G_{IJ}$ to be a flat metric over the $D$ dimensional spacetime in which the string lives.
Going through the motions one then obtains a beta function for $d \to 2$ given by
\beq \label{beta2}
\beta_G = - (26-D)\cdot \frac{2}{3} G^2 \,.
\eeq 
The difference between \eq{betaep} and \eq{beta2} arises from the lack of conformal fluctuations contributing to the later which therefore do not cancel the Jacobian $J_0$ which in turn  leads to an additional term in the flow equation along with a contribution from the $D$ scalar fields.
The beta function \eq{beta2} is the expected result reflecting the fact that when $26$ scalar fields are added to the action the conformal anomaly vanishes. 
Thus we reproduce the result that the bosonic string lives in $D=26$ dimensions, in which case we do not need to bother about the conformal anomaly \cite{Polyakov:1981rd}.
If $D<26$ the beta function \eq{beta2} does not vanish and the conformally anomaly should be taken into account  by including the induced action \eq{Sanom}.
This has been investigated recently in \cite{Codello:2014wfa} where it was found that in $d=2$ the ghosts alone gives the ``$-26$'' contribution found whereas here the ``-26'' arises from a combination of ghosts and gauge fixing.  
 This is nonetheless consistent with our result \eq{beta2} since the conformal fluctuations which are present there essentially play the role of one of the $D$ scalar fields. 

 We conclude that \eq{betaep} is the physically meaningful gauge independent beta function for $\epsilon>0$ in agreement with the physical beta function defined by \eq{physicalbeta}. On the otherhand in strictly $d=2$ one needs to take into account the conformal invariance (or anomalous breaking thereof)  to correctly evaluate the functional integral and find the corresponding beta function \eq{beta2}. The difference between \eq{betaep} and \eq{betaepex} arises since \eq{betaep} is the result of quantising general relativity in $d >2$ where conformal fluctuations do not propagate and are exactly canceled by the Jacobian $J_0$. The beta function \eq{betaepex} arises when working in the limit $d \to 2$ where, due to the particular choice of gauge/parameterisation, the conformal fluctuations mimic the effect of the induced action \eq{Sanom} such that the cancelation with $J_0$ is no longer exact.

\subsection{ $d$-dimensional beta function for Newton's constant}
In $d > 2$ dimensions the integral $\mathcal{I}_{d/2-1}[C]$ depends on the regulator function. This must be the case since Newton's coupling $G_k$ is dimensionful and thus $G$ will always depend on the regulator at least up to rescalings of $k$. However it is interesting to note that for an optimised cutoff \cite{Litim:2001up} of the form $ C_{\rm opt} = (z_0- z) \Theta(z_0-z)$, where  $\Theta(x)$ is a Heaviside theta function and $z_0$ is a positive constant, one obtains
\beq
\mathcal{I}_{d/2-1}[C_{\rm opt}] = z_0^{\frac{d-2}{2}} \frac{2}{d -2} \,,
\eeq
valid in all dimensions $d>2$. Hence a property of the optimised cutoff is to set all higher order corrections in \eq{Idto2} to zero up to an arbitrary rescaling.
For $z_0^{\frac{d-2}{2}} =(4 \pi)^{\frac{d}{2}-1} \Gamma(\frac{d}{2})$ one then obtains the beta function
\beq \label{betanorm}
\beta_G = (d-2) G - \frac{2}{3}( 18- N_g) \, \, G^2 \,,
\eeq 
where only the universal factor of $b$ remains.
This form of this beta function is clearly universal up to the normalisation of the RG scale $k$ or equivalently $G$. Thus although $G$ itself is not a physical observable it is clear that the sign of the quantum correction to $\beta_G$ is universal since one has $\mathcal{I}_n[C]>0$ for all regulator functions $C$.

For $N_g < 18$ the beta function implies $G_k$ decreases as the scale $k$ is increased. The running of $G$ stops at a UV fixed point given by,  
\beq
G_* = \frac{d-2}{2}  \cdot \frac{3  (4 \pi)^{\frac{d}{2}-1} \Gamma\left(\frac{d-2}{2}\right)}{18-N_g } \frac{1}{\mathcal{I}_{d/2 -1}[C]}  \,.
\eeq
This fixed point exists for positive $G_*$ in all dimensions for which $N_g < 18$ and describes an asymptotically safe quantum field theory at high energies.
However since this is a semiclassical result this interpretation is rather premature and one would like to go beyond this approximation to confirm this conclusion.
Indeed the critical exponent 
\beq
1/\nu \equiv - \frac{ \partial \beta_G}{\partial G}(G_*) = d-2\,,
\eeq 
which describes how $G$ approaches $G_*$ is simply the canonical one, independent of the regulator. To compute quantum corrections to $\nu$ one must go beyond the semi-classical approximation by computing $\beta_G$ to higher orders in $G$. This will be investigated in a companion paper \cite{Falls1} where we exploit a non-perturbative approximation to find quantum corrections to $\nu$.

\subsection{Paramagnetic dominance} 
From the form of the beta function \eq{betanorm} we observe that the increase of spin two degrees of freedom is responsible for the loss of asymptotic safety in higher dimensions. The term $-18$ which allows for asymptotic safety in $d<8$ dimensions is universal and independent of the dimension. The origin of this term is the interactions of the metric fluctuations with the curvature of spacetime, and represents the `non-Abelian' nature of gravity, in agreement with the ideas put forward in \cite{Nink:2012vd}. In particular it's origin is the 'non-minimal' coupling terms $E_n$ of the differential operators \eq{deltaE}

Following \cite{Nink:2012vd} we then denote terms proportional to the non-minimal couplings $E_n$ as `paramagnetic' and those arising from the Laplacian as `diamagnetic'  to obtain the expression 
\bea
A_{2,n} - 2 A_{2,1}  &=&  \big(\underbrace{d(d+1)/2 - 2 d}_{\rm diamagnetic}-\underbrace{6(\tr_2 E_2 - 2 \tr_1 E_1)}_{\rm paramagnetic}\big)\,\,\, \frac{R}{6}  \\
  &=&\big(\underbrace{N_g}_{\rm diamagnetic}-\underbrace{18}_{\rm paramagnetic}\big)\,\,\, \frac{R}{6} \,.
\eea
This confirms the results of \cite{Nink:2012vd} in a gauge independent setting: the physical mechanism behind asymptotic safety is related to the metric fluctuation's paramagnetic interaction with the curvature of spacetime. This effect is countered in $d>3$ dimensions by the diamagnetic interactions encoded in the Laplacians which dominate starting in $d\geq 8$ dimensions.

\section{Renormalisation of the vacuum energy} \label{vacuum}
As we have seen the absence of the cosmological constant in the beta function for Newton's constant results from demanding gauge independence and therefore its presence in gauge dependent beta functions is most likely unphysical.
This can be seen as resulting from the fact that the cosmological constant is not a mass for the graviton. Rather the vacuum energy only couples to the conformal fluctuations $\sigma$ which are not themselves dynamical. Nonetheless the vacuum energy $\lambda = \frac{\bar{\lambda}}{8 \pi G}$ has also a $k$ dependence given by,
\beq \label{betalambda}
\partial_t \lambda = - d \lambda +\frac{N_g }{(4 \pi)^{\frac d 2} \Gamma(d/2) } \mathcal{I}_{\frac d 2}[C]
\eeq
The quantum correction being proportional to $N_g$.
This beta function has a fixed point, 
\beq
\lambda_* = \frac{1}{d}  \frac{N_g }{(4 \pi)^{\frac d 2} \Gamma(d/2) } \mathcal{I}_{\frac d 2}[C]\,.
\eeq 
Note that in the absence of any propagating degrees of freedom, i.e. when taking $d=3$, the quantum corrections to this beta function vanish. This result is just related to the divergencies of the vacuum encountered in any quantum field theory in flat spacetime. In this case we are just looking at the graviton contribution to the vacuum energy. To see this lets write  \eq{betalambda} explicitly in terms of the IR cut off scale $k$ and the dimensionful vacuum energy $\lambda_k = k^d \lambda$, then one has
\beq
\partial_t \lambda_k =d k^d \lambda_* \,,
\eeq 
introducing the bare UV scale $\Lambda$ defined by $\lambda_{k=\Lambda} \equiv \lambda_b$ we can solve this equation to find,
\beq
\lambda_k = \left(k^d  - \Lambda^d \right)  \lambda_*  + \lambda_b \,.
\eeq
In the limit $k \to 0$ we remove the IR regulator to obtain the observed vacuum energy $\lambda_{0} \equiv \lambda_{k= 0}$. In turn we may express the bare vacuum energy in terms of $\lambda_{0}$ as
\beq
\lambda_b = \lambda_{0} +  \Lambda^d \lambda_* \,.
\eeq 
Thus the `beta function' \eq{betalambda} is nothing but the statement that we must include a counter term proportional to the number of local degrees of freedom in order to get a finite renormalised vacuum energy $\lambda_{0}$. 

\section{Continuum limit}  \label{continuum}

As discussed previously the existence of a continuum limit, $\Lambda \to \infty$, relies on the sign of $b$, the $G^2$ coefficient of $\beta_G$. To understand this we note that the sign of the bare Newtons constant $G_b$ must be positive if the functional integral is to make sense. If one is forced to take $G_b<0$ to renormalise the theory one must concede that the theory cannot be fundamental and that the continuum limit does not exist.  From the \eq{betaG} one may infer the energy dependence of the Planck mass $\kappa_k = \frac{1}{16 \pi G_k}$,
\beq
\kappa_k = (k^{d-2} - \Lambda^{d-2}) \kappa_* + \kappa_b  \,,
\eeq
where $\kappa_* = \frac{1}{16 \pi G_*}$ is the regulator dependent value of the fixed point for the dimensionless quantity $\kappa = k^{2-d} \kappa_k$.
Taking $k \to 0$ one has the renormalised Planck mass $M_{\rm Pl}^{d-2} \equiv \frac{1}{16 \pi G_N} =  \kappa_{k=0}$ where $G_N$ is the measured value of Newtons constant. Then we have
\beq
\frac{1}{16 \pi G_b} = \frac{1}{16 \pi G_N} + \Lambda^{d-2} \kappa_* \,.
\eeq
 for the functional integral to make sense one must have $G_b >0$ otherwise the kinetic terms will have the wrong sign and the theory will breakdown. 
 In particular for the semiclassical theory to be valid we need
\beq
\frac{1}{16 \pi G_N }  > - \Lambda^{d-2} \kappa_* \,.
\eeq
Now we know that $\kappa_* \propto (18 - N_g)$ and that the proportionality constant is positive from which we obtain the inequality for $N_g > 18$
\beq \label{UVcutoffbound}
\left(\frac{\Lambda}{M_{\rm Pl}}\right)^{d-2} <16 \pi |G_*| \,.
\eeq
Thus, while for $18 < N_g$ there is no maximum value for which the UV scale can take, for $N_g > 18$ \eq{UVcutoffbound} provides a bound.
In the latter case we can say that the semiclassical theory predicts its own downfall at a finite energy scale. In the former case no bound exists and the semiclassical approximation suggests that the theory is asymptotically safe.

\section{Inclusion of matter} \label{matter}  
It is rather straight forward to include the effects of matter. This is done by adding a matter action to the bare Lagrangian and adding a corresponding regulator term (see for example \cite{Dona:2013qba}) or by computing the regulated one loop effective action \cite{Calmet:2008tn}. Doing the former adds new traces to the right hand side of \eq{oneloopflow} proportional to the number of matter fields. In turn these give contributions to the beta functions. Here we give the result in $d=4$ spacetime dimensions, although the result is easily generalised to arbitrary $d$, \footnote{Here we assume the fields are massless and minimally coupled to gravity. The beta function will be modified if these assumptions are dropped.}
\beq \label{betamatter}
\beta_G =2 G - \frac{2}{3}( 16- 2 N_D + 4 N_M - N_s) \, \, G^2 \,,
\eeq 
up to the normalisation factor, where $N_D$ is the number of Dirac fermions, $N_M$ is the number of gauge fields and $N_s$ is the number of scalars. Plugging in the values for the standard model ($N_D = 45/2, \,N_M=12,\,N_s = 4$) we have
\beq \label{betaSM}
\beta_G = (d-2) G - \frac{2}{3}\cdot 15 \, \, G^2 \,,
\eeq 
which continues to be anti-screening as in the pure gravity case. We therefore can conclude that the semiclassical theory of the standard model coupled to gravity is not predicting its own down fall at the quantum level and the existence of a continuum limit along the lines of asymptotic safety remains a possibility. In particular the beta function predicts asymptotic safety provided
\beq
16- 2 N_D + 4 N_M - N_s > 0
\eeq
otherwise $G_*$ is negative and the bound \eq{UVcutoffbound} applies.
We observe that by adding large numbers of fermions or scalars the theory can break down a fraction of the Planck scale. This then puts constraints on effective theories that contain a large number of scalars or fermions. 

Similarly the renormalisation of the vacuum energy at one-loop is given by
\beq \label{betalambda}
\partial_t \lambda = - d \lambda +\frac{N_{\rm Bose} - N_{\rm Fermi} }{(4 \pi)^{\frac d 2} \Gamma(d/2) } \mathcal{I}_{\frac d 2}[C]\,,
\eeq 
where the number of bosons $N_{\rm Bose}$ (including gravitons)  and the number of fermions $N_{\rm Fermi}$ enter with opposite signs.
This is the expected result and entails that the vacuum energy is not renormalised at one-loop for super-symmetric theories.

\newpage
 \section{Conclusions} \label{conclude}

In this paper we have derived a gauge independent effective action for quantum gravity at one-loop which only depends on physical fluctuations of the metric.
To achieve gauge independence we have used a specific parameterisation of the metric for which the volume element is linear in the conformal fluctuation $\sigma$. 
We have then used renormalisation group techniques to compute the one-loop beta function for Newton's constant.  When introducing a regulator scheme we have been careful to do 
so in such a manner that cancellations, which should occur when 
diffeomorphism invariance is present, continue to occur in the presence of the regulator. In particular this relies on choosing a type II regulator of the form \eq{RegChoice} while 
choosing the same regulator function $C(z)$ for the different fields.  Different choices can lead to unphysical results. In particular the absence of the local fluctuations in $d=3$ may not 
be manifest if the regulator functions are chosen differently.

Here we offer an interpretation of the gauge independence of our results. A key point is that our calculations are performed on Einstein spaces for which the trace-free Einstein equations are solved. Thus we are on-shell with respect to all but one of the equations of motion, this one being essentially the equation of motion for the volume element. Now, one way to ensure gauge independence in any gauge theory is to work in terms of physical quantities rather than gauge variant fields such as the metric \cite{Vilkovisky:1984st}. Since we single out the volume element by our parameterisation we are therefore singling out the corresponding physical quantity which in this case is the gauge invariant field $\phi$ given in \eq{DBredef2}. Thus we solve the problem of finding a physical parameterisation on an Einstein space (at least at one-loop order). This is seen explicitly since without the gauge fixing terms the quadratic action involving $S^{(2)}$ is independent of both $\xi$ and $\psi$ and hence diffeomorphism invariant. 

A further insight is  gained by noting that the effective field equations for the field $\sigma$ are solved for the expectation value $ \langle \sigma \rangle$ which then gives direct access to the mean space-time volume 
\beq
\langle V(\gamma) \rangle = V(g) +  \int d^dx  \sqrt{g} \,\frac{\langle \sigma \rangle}{2} \,.
\eeq 
This can be considered as advantage since to compute $\langle V(\gamma) \rangle$ for the standard effective action involves an infinite number of terms consisting of $n$-point functions to all orders by expanding the volume element $\sqrt{\gamma}$ in $h_{\mu\nu}$. 
We therefore believe that  the general scheme put forward here is physically well motivated and should be used in approximations 
beyond the simple one employed here. The fact that the universal factor \eq{betaintro} only depends on the dimensionality of spacetime through the number polarisations $N_g$ would 
seem to be a vindication of this claim and that it represents the analogy to the factor $11 - \frac{2}{3} N_f $ in QCD.

\section*{Acknowledgements}
The author would like to thank Christof Wetterich, Jan Pawlowski and Roberto Percacci for insightful discussions on this work. This work was supported by ERC-AdG-290623.

 \bibliography{ASreferences2,myrefs}

 \end{document}